\documentclass[aps,prl,preprint,groupedaddress]{revtex4-1}

\usepackage{picinpar,graphicx}
\usepackage{wrapfig}
\usepackage{caption}
\usepackage[letterpaper,left=2.8cm,right=2.6cm,top=2.4cm,bottom=2.6cm]{geometry}
\usepackage{parskip}
\usepackage{amsmath}
\usepackage{cases}

\begin{document}


\title{Investigation of a parametric instability between 
ELF and VLF modes driven by antennas immersed in a cold, magnetized plasma}


\author{D. Main}
\email[]{daniel.main@gmail.com}
\affiliation{T2Sys Inc., Beavercreek, Ohio 45431, USA}
\author{V. Sotnikov}
\author{J. Caplinger}
\affiliation{Air Force Research Laboratory (AFRL/RY), Wright Patterson AFB, Ohio, 45433, USA}
\author{D. V. Rose}
\affiliation{Voss Scientific, Albuquerque, New Mexico 87108}


\date{\today}

\begin{abstract}
We have studied the behavior of a VLF, ELF and combined ELF/VLF antenna immersed in a cold, 
magnetized plasma using a fully kinetic, three dimensional Particle-in-Cell simulation code
called Large Scale Plasma (LSP). All the antennas are modeled as magnetic dipoles ($\rho_{ant}=0$) and are 
assigned a time varying current density within a finite sized current loop. 
The VLF antenna is driven at 10 Amps with a frequency ($\omega_{VLF}$) greater than the lower 
hybrid frequency ($\omega_{LH}$), while the ELF antenna is driven at 3 Amps with a frequency 
($\omega_{ELF}$) less than $\omega_{LH}$.  The combined ELF/VLF antenna (which we call a 
“parametric antenna”) includes both antennas driven simultaneously in 
the same simulation domain. We show that the parametric antenna non-linearly 
excites electromagnetic (EM) Whistler waves to a greater extent than the VLF antenna 
alone. We also show that the parametric excitation of EM Whistler waves leads to greater
emitted EM power (measured in Watts) compared with a VLF antenna alone. 
\end{abstract}

\pacs{}

\maketitle

\section{}
Whistler waves are ubiquitous in the space environment and have been observed in the magnetotail
\cite{VKV14}, ionosphere \cite{MS16}, solar wind \cite{GSP94}, 
other planets \cite{HMH95,OR95} and numerous laboratory experiments \cite{Stenzel99}.  
Some of the earliest observations of whistler waves were correlated with lightning strikes in which
the whistler wave is guided by an ionospheric duct \cite{Smith1961}. An important 
aspect of whistler waves is that they are known to cause pitch angle scattering of highly 
energetic electrons, for example, in Earth's radiation belt \cite{MAM16}.  Furthermore, 
EM whistler waves weakly decay from their source region and 
travel great distances along the background magnetic field. The group velocity 
(and hence the energy) of the EM whistler travels in a cone with a peak angle of $19.5^o$ w.r.t. 
Earth's magnetic field, which is known as the shadow boundary and is determined by the long
wavelength inflection point in the dispersion relation \cite{FKS87}. 
Therefore, as whistler waves propagate great distances along
Earth's magnetic field, they carry with it energy that pitch angle scatter highly 
energetic particles, causing these particles to violate the frozen in condition.

One method for generating whistler waves in a cold, magnetized plasma is 
with a magnetic or electric loop antenna driven in the frequency range
$\omega_{LH} < \omega_{VLF} \ll \omega_{ce}$ \cite{FG71,WB72},which we call a VLF antenna
and $\omega_{ce}$ is the electron cyclotron frequency.
In a magnetic loop antenna \cite{Karpman86} the charge density in the antenna ($\rho_{ant}$) 
equals 0 and the current density varies with time at frequency $\omega$. 
In an electric loop antenna $\rho_{ant} \not= 0$ and the charge density varies 
spatially with frequency $\omega$. It has been shown \cite{Karpman86} that the two are 
equivalent and both result in singularities in the electric field within a 
cone of angle $\theta_c$ measured off the magnetic field direction, 
though the electric field singularity is stronger in the electric loop antenna. 
In a plasma with no dissipation, the resonance cones form at an angle given by
\begin{equation}\label{cone_angle}
sin^2\left(\theta_c\right) = \frac{\omega^2_{VLF}\left(\omega^2_{pe}+\omega^2_{ce}-\omega^2_{VLF}\right)}
                                  {\omega^2_{pe}\omega^2_{ce}}
\end{equation}

In a plasma with dissipation, the singularities become finite within the angle $\theta_c$ 
in a spatially localized resonance cone. As noted in previous work, much of 
the source power due to a VLF antenna is radiated as electrostatic Lower Oblique 
Resonance (LOR) modes [also referred to as quasi-electrostatic whistler waves] 
which decay as $R^{-1}$ ($R$ is the distance from the antenna) 
away from the source antenna, whereas the EM whistler wave decays as $R^{-1/2}$ \cite{FKS87}. 
Considerable experimental work has shown that the loop antennas driven 
within the frequency range $\omega_{LH} < \omega \ll \omega_{ce}$ 
form LOR waves as expected \cite{US14}. However, in these efforts, 
it is not clear how much of the power is radiated as EM whistler waves compared 
with the LOR modes. 

One method that has been proposed to increase the wave power in the EM
whistler wave is through a parametric interaction between LOR modes and a low
frequency density perturbation generated by a dipole antenna which excites
ion sound waves \cite{FKS87,SSA93}.
In this paper, we generate a low frequency density perturbation with a loop
antenna and excite ELF waves instead ion sound waves. We call the low frequency
loop antenna an ELF antenna which is driven at a frequency $\omega_{ELF} < \omega_{LH}$,
and show below that the ELF antenna drives a fast magnetosonic wave which
causes the low frequency density perturbation. We call an antenna consisting of a 
combined ELF and VLF antenna 
(occupying the same volume but driven at two
different frequencies) a parametic antenna.
We demonstrate in this paper an increase in the EM whistler wave power in
the parametric antenna simulation compared with the simulation of a VLF 
antenna alone and attribute
this increase in wave power to a parametric interaction 
between the LOR modes and a low frequency density perturbation.  
We find in the parametric antenna that 
whistler waves are excited on combination frequencies $\omega_{VLF} \pm \omega_{ELF}$, 
as expected from theoretical work. 

We now describe the 3D simulation set up for three different antennas immersed in a magnetized plasma. 
We have performed three different fully kinetic simulations which we call Run 1, Run 2, and Run 3.
Run 1 contains the ELF antenna, Run 2 contains the VLF antenna, and Run 3 contains the parametric 
antenna. All three antennas are identical except the frequency at which they are driven. 
Besides the different antennas, all three runs are identical.
We present results that demonstrate the formation of resonance cones at angles consistent with 
theory and the non-linear excitation of EM whistler waves in the parametric antenna simulation. 
Furthermore, we will compare wave spectra with linear theory to demonstrate that the wave structures
that form in the three separate simulations are consistent with theory.
The simulation domain is established in a Cartesian volume such that -600 m $<$ $x,y$ $<$ 600 m and 
-750 m $<$ $z$ $<$ 750 m, where $x$ and $y$ are perpendicular to the external magnetic field and z is parallel. 
The number of cells used is $n_x=n_y$ = 600 and $n_z$ = 750 so that the grid size in all dimensions is 2 m. 
The spatial grid size was chosen based on trial 2D simulations in which we varied
the grid size and compared the evolved field structures. In these different runs, we set the grid size
to 0.50 m, 1 m, 2 m and 4 m. We find good agreement with linear theory
in the different 2D runs up to a grid size of 2 m. 
Therefore, for the 3D runs, we chose to use the 2 m grid size due to computational constraints. 
For these simulation, we use 8 plasma particles per cell (4 electrons, 4 ions), for a total 
of $\sim 2.2 \times 10^9$ particles. The mass ratio of ions to electrons is 1836:1, so that these simulations 
are assuming the ion species is hydrogen.  We use an implicit, energy conserving algorithm to 
provide the Lorentz force particle push and also to solve for the self-consistent EM fields 
\cite{WRC04,WRC06}. The time step (dt) used is 3 times the CFL limited time step, which equates to 
a value of $dt \approx 11$ ns. This time step was chosen by varying it from 1 to 5 times the 
CFL limited time step in otherwise equivalent 2D simulations. 
Above 3 times the CFL limited timestep, noticable difference were observed
in the evolved field structures.      

The plasma parameters for the simulation results presented in this paper are the following: 
The background electron and ion density is $10^5$ cm$^{-3}$. The background magnetic field points 
in the z-direction and is 0.30 Gauss. We assume hydrogen ions. The temperature of the plasma is 
set to 0 which reduces the numerical noise in the simulation. We have performed simulations 
with a finite temperature and achieve similar results to the ones presented here. Furthermore, 
by setting the plasma temperature to 0, we do not need to replace particles that can leave the 
simulation through the outlet boundaries (discussed in the next paragraph).  We have set 
$\omega_{VLF}$ = 1.31$\times 10^6$ rad/s $\approx$ 11$\omega_{LH}$ and $\omega_{ELF}$ = 1.04$\times10^5$ rad/s
$\approx 0.88\omega_{LH}$. 

The simulation uses outlet boundaries \cite{BL05} which attempt to match the outgoing plasma
waves in the simulation domain with a virtual wave that forms outside the simulation domain. 
This matching condition allows the wave to propagate out of the simulation domain and minimizes 
reflections back into the simulation domain.  We construct a “loop” antenna in the center of the 
simulation domain using a “volume” model in LSP. This volume model generates a uniform dipole 
current with a set frequency within a fixed volume of space and therefore, via Maxwell's equation,
generates a dipole electric field. However, the volume model only allows 
us to define a dipole electric field in a solid region. To construct a dipole loop, we glue four 
solid volumes together with each volume composing a side of a cube loop (this is analogous to 
a square donut). At the four corners, we superimpose conducting volumes which we find reproduces 
better the Lower Oblique Resonance cones. The antenna is placed in the center of the simulation domain
such that the normal vector to the loop antenna lies in the $z$-direction (which is also the direction of
the external magnetic field). Therefore, the plane of the loop lies in the $x-y$ plane. The square antenna has an
inner width (which represents the hollow portion) of 700 cm and and outer width of 1100 cm so that the thickness
of the antenna is 400 cm. In the $z$-direction, the thickness is 1100 cm. 

\begin{figure}[h]
  \begin{center}
    \includegraphics[scale=0.5]{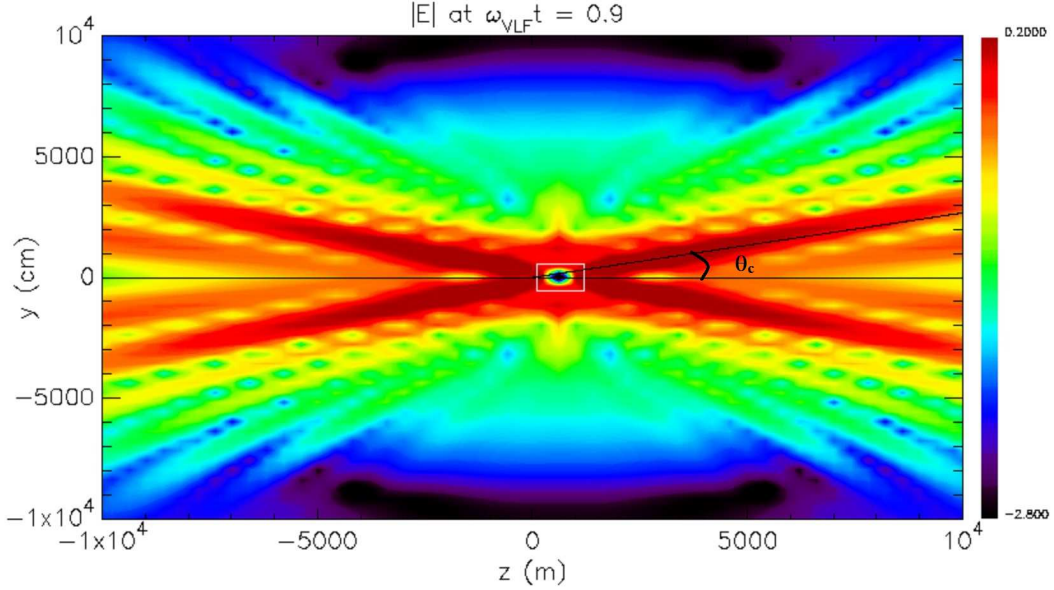}
  \end{center}
  \caption{Magnitude of the electric field early in the parametric antenna simulation 
showing the formation of the LOR cones. The black square represents the antenna.  Note that only a 
portion of the simulation domain in shown in this Figure. The magnetic field is in the z-direction.}
\label{LOR_Cones}
\end{figure}

In Figure \ref{LOR_Cones} we show a 2D slice of the LOR cones. The angle that these structures form is consistent 
with theory \cite{FG71} and is found to be $\sim 15^o$ for the plasma parameters that we have used.  
We varied $\omega_{VLF}$ in 2D simulations (in the x-z plane) and we find the resonance cones form at a 
smaller angle when $\omega_{VLF}$ decreases, consistent with Equation \ref{cone_angle}.  
Therefore, it is clear that 
we can reproduce the resonance cones discussed in previous experimental and theoretical results.  
However, because these waves 
have a large electrostatic component, the  $\vec{J} \cdot \vec{E}$ power generated 
does not propagate far from the antenna. The goal of 
this paper is to demonstrate, using a fully kinetic PIC model that we can parametrically couple 
the electrostatic LOR waves with the electromagnetic magnetosonic waves and pump additional 
power into the electromagnetic whistler waves. 

Prior research \cite{FKS87,SSA93} has discussed in greater depth the dispersion curve for the
EM whistler and ES Lower Oblique modes. As discussed in these papers for values of
$k_{\perp} \ll \omega_{pe}/c$ ($c$ is the speed of light and $k_{\perp}$ is the wave vector 
perpendicular to the external magnetic field), the wave is electromagnetic and for 
$k_{\perp} \gg \frac{\omega_{pe}}{c}$ the wave is electrostatic. 
Therefore, the LOR waves have values of $k_{\perp} \gg \omega_{pe}/c$. 
The strict upper limit on the value of $k_{\perp}$  above which the wave
is quasi-electrostatic is known as the shadow boundary which
is the long wavelength inflection point of the refractive index surface.
However, this corresponds to a value of  $k_{\perp} \approx 0.1 \omega_{pe}/c$,
and therefore the purely EM whistler mode resides in the range
$0 < k_{\perp} \lesssim 0.1 \omega_{pe}/c$, which requires exceedingly large
computational domains to resolve.
Furthermore, according to Equations (4) and (5) from \textit{Fiala et. al.} \cite{FKS87},
the ES portion of the disperion curve is linear in $k_z$. We find that the linear
portion of the disperion curve starts near $k_{\perp} = \omega_{pe}/c$.
Therefore, in this paper, we define waves with $k_{\perp} < \omega_{pe}/c$ to be
EM and $k_{\perp} > \omega_{pe}/c$ to be ES, which is obviously an approximation,
but a necessary one due to our limited computing resources. 
To calculate the EM wave power (measured in Watts), we have developed a 
“k-space filter” which allows us to calculate the EM contribution to the
electric fields ($\vec{E}$) and current density ($\vec{J}$).
Essentially, we calculate the Fourier Transform (FT) of the fields such that
$\textrm{FT}\left\{\vec{E}(x,y,z)\right\}=\mathcal{\vec{E}}(k_x,k_y,k_z)$ and
$\textrm{FT}\left\{\vec{J}(x,y,z)\right\}=\mathcal{\vec{J}}(k_x,k_y,k_z)$. Next
we set $k^2_{\perp}=k^2_x+k^2_y$ and invoke a filter in $k$-space according to the following
equations:
\begin{align} \mathcal{\vec{E}}_{EM}(\vec{k}) =
\begin{cases}
  \mathcal{\vec{E}}(\vec{k}) & \text{if $k_{\perp} < \frac{\omega_{ce}}{c}$}  \\
                   0         & \text{if $k_{\perp} > \frac{\omega_{ce}}{c}$}  
\end{cases}
\label{filtera}
\end{align}
\begin{align} \mathcal{\vec{E}}_{ES}(\vec{k}) = 
\begin{cases} 
  \mathcal{\vec{E}}(\vec{k}) & \text{if $k_{\perp} > \frac{\omega_{ce}}{c}$}  \\
                   0         & \text{if $k_{\perp} < \frac{\omega_{ce}}{c}$}
\end{cases}  
\label{filterb}
\end{align}             
Where the subscript EM/ES denotes electromagnetic and electrostatic portion of the field. 
The same filter is applied to the self-consistent current density, $\mathcal{\vec{J}}$.  Note
that in Equations \ref{filtera} and \ref{filterb}, all values of $k_{\|}$ are included in the EM and ES portions of the
E- and J-fields. We next compute the power due to the EM portion of the fields, which we demonstrate below is 
mainly due to the EM Whistler wave based on the good agreement between the Whistler disperion and the Fourier 
spectrum of the electric field. 
We use the following equation to compute the power:
\begin{equation}
P_{EM}=\frac{1}{2}\int\mathcal{\vec{E}}_{EM}\cdot\mathcal{\vec{J}}^*_{EM}+\mathcal{\vec{E}}^*_{EM}\cdot\mathcal{\vec{J}}_{EM}d^3k
\label{EM_power}
\end{equation}
Where the * indicates complex conjugate. In order to compare Equation \ref{EM_power} with the theoretical output of the antenna, 
we have also computed the $\vec{J}\cdot\vec{E}$ power from the antenna in the following way:
\begin{equation}
P_{Ant}=\frac{1}{2}\int\mathcal{\vec{E}}_{EM}\cdot\mathcal{\vec{J}}^*_{Ant}+\mathcal{\vec{E}}^*_{EM}\cdot\mathcal{\vec{J}}_{Ant}d^3k
\label{Ant_power}
\end{equation}
where $\mathcal{\vec{J}}_{Ant}$ is the FT of the currrent density in the antenna. 
Of course, Equations \ref{EM_power} and \ref{Ant_power} are nearly identical. Essentially, 
in using Equation \ref{EM_power}, we are only considering the power which is non-linearly pumped
into the EM fields by the plasma currents which act as a large antenna driven by the parametric
instability. 
The lower bound on the integrals in Equations \ref{EM_power} and \ref{Ant_power} is limited by the 
perpendicular box size, which is 1200 m in 
our simulation. Therefore, the lower bound is $\sim 0.08 \omega_{pe}/c$ and the upper bound is $\omega_{pe}/c$
due to Equation \ref{filtera}

\begin{figure}[h]
  \begin{center}
    \includegraphics[scale=.75]{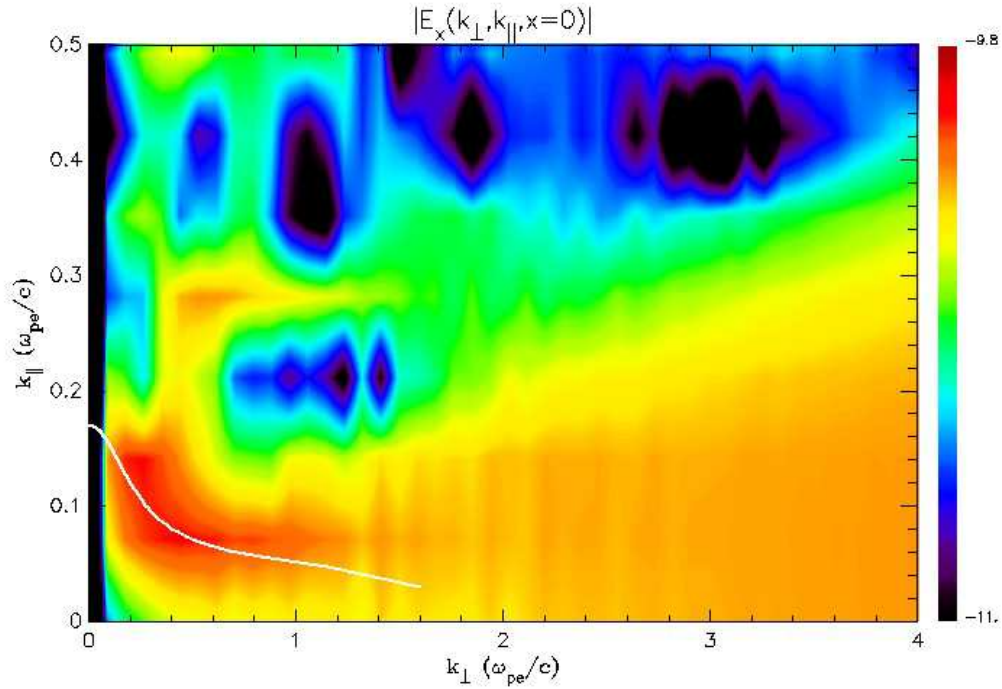}
  \end{center}
  \caption{The wave power calculated from $E_x$ through the x=0 plane from the 3 Amp ELF simulation at $\sim$ 0.20 
           ELF periods. The white curve represents the solution to the FM disperion relation.}
\label{ELF_dispersion}
\end{figure}

We have performed two simulations with only ELF antennas, one driven at 1 Amp and the other driven at 3 Amps. 
We have also performed a simulation with only a VLF antenna driven at 10 Amps. Finally, we performed two simulations 
with a parametric antenna such that the 
ELF/VLF currents are 1 Amp/10 Amps and 3 Amps/10 Amps. We find little difference between the 1 Amp/10 Amp parameteric
antenna simulation and the 10 Amp VLF antenna simulation, indicating that there is little non-linear interaction 
between the LOR modes and density perburbations driven by the ELF antenna in this case. 
Therefore, all results 
discussed in the remainder of this paper are for the 3 Amp/10 Amp parametric antenna simulation.  
We expect the ELF antenna to drive Fast Magnetsonic 
(FM) waves which has a disperion equation given by Equation (3) in \textit{Sagdeev et. al.} \cite{SSS77}.
We show in Figure \ref{ELF_dispersion} the $k$-space spectra from the ELF simulation with the white curve
representing the solution to the FM disperion in \textit{Sagdeev et. al.} \cite{SSS77}. 
The good fit between the linear disperion curve and the power spectra from the simulation
indicates that we are resolving the necessary wave numbers to drive the FM mode. 

\begin{figure}[h]
  \begin{center}
    \includegraphics[scale=.5]{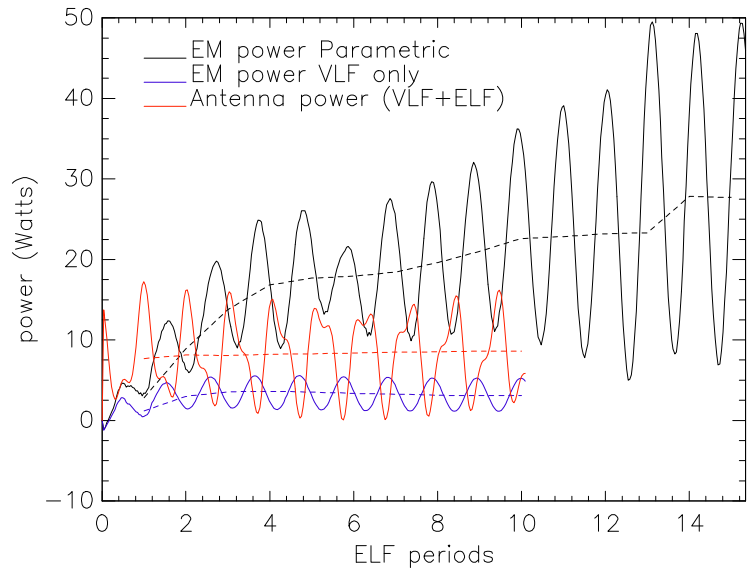}
  \end{center}
  \caption{EM power calculation (blue and black curves) using Equation \ref{EM_power} and 
          the power generated by the antenna (red) using Equation \ref{Ant_power}. The dashed curve
          represents the average over one ELF period.}
\label{power}
\end{figure}

For comparison, the results of the two calculations from Equations \ref{EM_power} and \ref{Ant_power} 
are shown in Figure \ref{power}.  
We saved the full 3D electric field, magnetic field, and current density every 200 time steps, which allows 
us to resolve 2 data points per VLF period and 26 data points per ELF period. We ran the parametric simulation for 
about 80000 time steps, which is about 15 ELF periods.  The simulations were performed on massively parallel 
computing clusters using several thousand processors and took about 15 days. 
To compare the parametric run with the 10 Amp/3 Amp antenna, 
we also ran a VLF simulation (driven at 10 Amps) and an ELF simulation (driven at 3 Amps) with identical 
simulation domain sizes, grid sizes and time steps.  This has allowed us to compare the linear and 
non-linear evolution of the plasma. Because the calculation from the VLF and
ELF simulations show that the power output from the antenna level off more quickly than the parametric 
antenna simulation, we ran
these two simulation for 50000 time steps. We note an oscillatory trend to all three data
sets, with a frequency of $\sim \omega_{ELF}$. However, given that the VLF simulation is 
independent of the ELF frequency, we surmise that we are driving a fundamental FM
mode in the plasma nearly independent of the ELF driving frequency. We tested this idea
by performing another parametric simulation in which the ELF antenna is driven at $\sim$ 0.52$\omega_{LH}$) 
(i.e. $\sim$60\% of the original frequency) and indeed find the same periodicity in the power calculation,
demonstrating that the fundamental frequency is independent of the driving frequency. 
We have averaged over 1 ELF period to smooth the oscillation and demonstrate an increase
in average power in the parametric antenna. This is shown as the dashed curves in Figure 
\ref{power}. The red curves in Figure \ref{power} represents the superposition of the linear power
generated by the ELF and VLF antennas run independently. 
In comparing the power generated by the superposition of the ELF and VLF antennas and the parametric antenna,
we note a factor of 3 increase in power. 
Not all the power generated by the antenna radiates away from the antenna. Some of the power 
generates ES waves and some of it heats the plasma, neither of which are accounted for in 
Equations \ref{EM_power} and \ref{Ant_power}. We have calculated the EM wave power from the VLF 
simulation alone using Equation \ref{EM_power} and also show this calculation as the blue curve
in Figure \ref{power} and note a factor of 7 gain between the VLF and parametric antennas. 

To demonstrate that the increase in power observed in Figure \ref{power} is in fact due to EM whistler 
waves, 
we show wave spectra (FT of field components) in the $y-z$ plane from the 10 Amp VLF simulation, 
and the 10 Amp/3 Amp parametric simulation. Based on previously developed theories \cite{FKS87,SSA93}, we infer
that the best explanation for the observed increase in the EM whistler wave power is due to a
parametric interaction between the FM wave and the LOR resonance waves. The whistler dispersion from 
\textit{Fiala et. al.} \cite{FKS87} is compared with wave power from the VLF and parametric simulations in 
all four panels. Panels (a) and (b) represent the FT of $E_x$ in the y-z plane at x = 40 m and
Panels (c) and (d) represent the FT of $E_z$ in the y-z plane at x = 400 m.  
The two left panels are from the VLF simulation and the two right panels are from the parametric simulation.
According to \cite{FKS87}, for $k_{\perp} << \omega_{pe}/c$, 
the wave corresponds to the EM whistler wave and for $k_{\perp} >> \omega_{pe}/c$, the wave corresponds to the LOR. 
Therefore, if we are exciting the EM whistler in the parametric antenna simulation, then we expect 
to observe smaller wave numbers excited. In comparing the VLF (left two panels) and parametric simulations 
(right two panels)
we observe that both follow the whistler dispersion curve well. However, in the parametric simulation, 
we also see that lower wave number modes have a greater wave power compared with the VLF antenna alone.
Close to the antenna at x=40 m [the antenna is placed close to the center of the simulation domain which
is at the coordinates (0,0,0)], we see large EM wave power in both the VLF and parametric antenna. Though
difficult to see from the color scale, the average wave power in the parametric antenna is $\sim$2 times greater
than the VLF antenna. However, it is very obvious that far from the antenna at x=400 m, the waves are dominated
by the EM whistler wave and that the parametric antenna wave power is much larger than the VLF wave power 
($\sim$ 10 times greater).

\begin{figure}[h]
  \begin{center}
    \includegraphics[scale=.4]{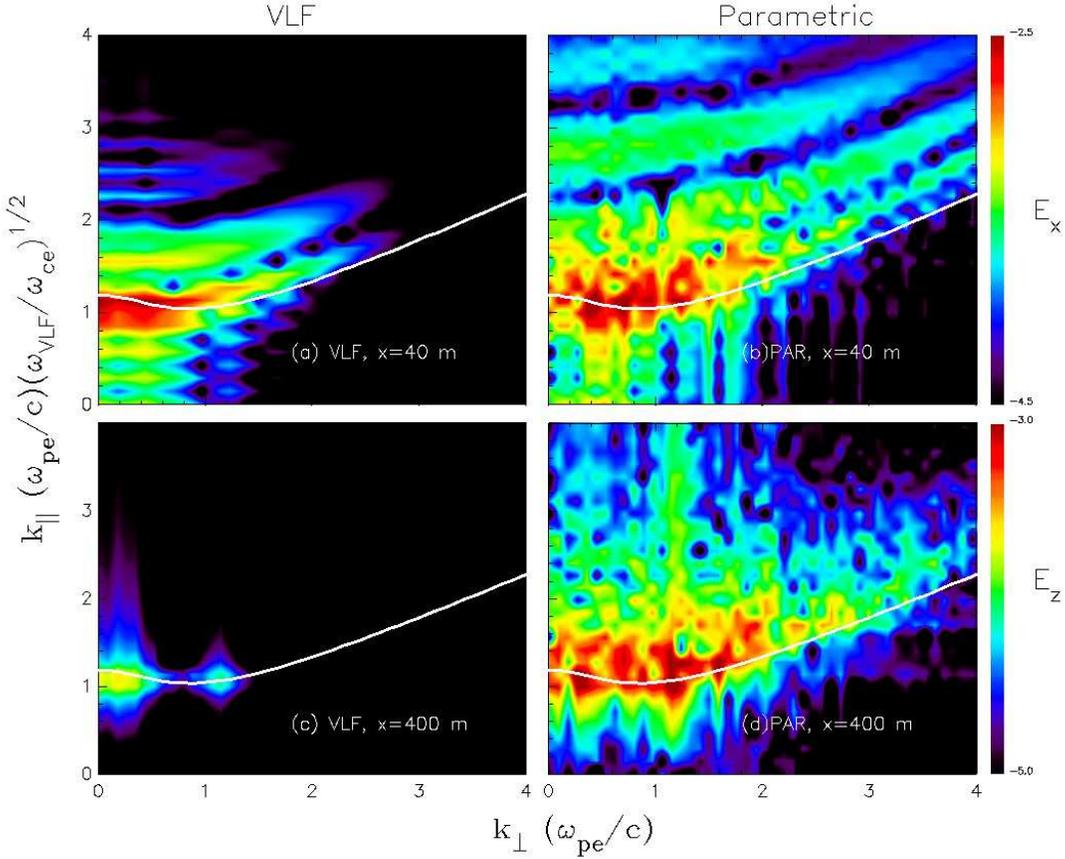}
  \end{center}
  \caption{(a,b) Wave power from the VLF (panel a) and parametric (panel b) simulations computed from $E_x$
           at x=40 m. (c,d) wave power from $E_z$ at x=400 m from the VLF (panel c) and parametric (panel (d)
           simulation. All data are from time step 46200 ($\sim$ 9 ELF periods).  }
\label{spectra}
\end{figure}

\begin{figure}[h]
  \begin{center}
    \includegraphics[scale=.4]{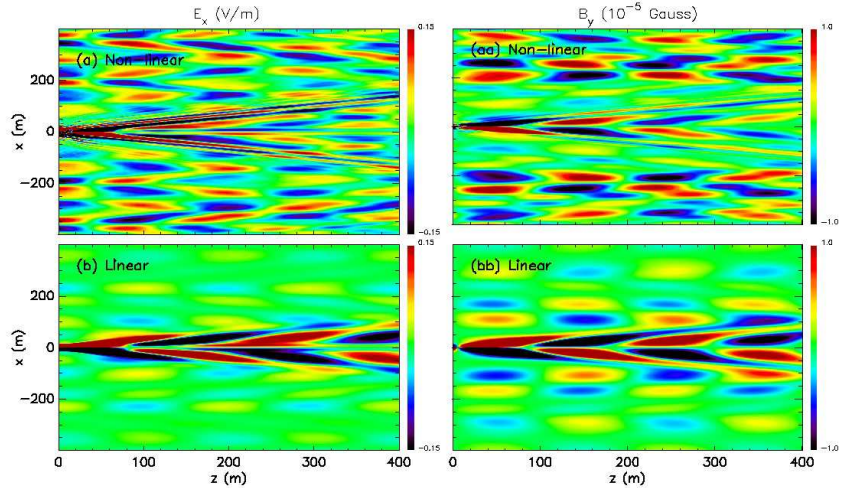}
  \end{center}
  \caption{(a,aa) Non-linear electric and magnetic field from the paramatric antenna simulation 
           as defined in Equation \ref{linear_NL}. (b,bb) Linear electric and magnetic fields also defined in Equation \ref{linear_NL}.
           All panels are plotted at time step 46200 ($\sim$ 9 ELF periods).}
\label{fig:linear_NL}
\end{figure}

Finally, we compare the non-linear wave amplitudes excited in the parametric simulation with 
the linear wave amplitudes. We define the linear fields as the superposition of the fields 
from the separate ELF and VLF simulation, and the non-linear fields as the total field from 
the parametric simulation minus the fields from the separate ELF and VLF simulations:
equation:
\begin{equation}
\begin{aligned}
E_L &= E_{VLF}+E_{ELF}  \\
E_{NL} &= E_{par}-E_L
\end{aligned}
\label{linear_NL}
\end{equation}             
Where E denotes the electric field. The same definition can be written for all the field quantities. 
We show $E_x$ and $B_y$ for both the linear and non-linear fields in Figure \ref{fig:linear_NL}. Note that
only a portion of the simulation domain is plotted. Notice in both runs 
considerable wave activity along the external magnetic field on the 
same axis as the antenna at the center of the figures. 
However, in the NL fields, we notice off-axis waves that fill the simulation domain that are not 
present in the linear fields. We attribute these off-axis waves to the EM whistler waves excited 
by the parametric interaction between the LOR waves and the FM waves. This
interpretation is consistent with Figure \ref{spectra} which shows that far from the antenna there
is considerable EM whistler wave power in the parametric antenna compared with VLF antenna.   
Figure \ref{fig:linear_NL}a demonstrates that significant non-linear wave activity occurs in the parametric antenna 
simulation and Figure \ref{fig:linear_NL}aa shows that these waves have a strong magnetic
field component, consistent with interpretation that these off-axis waves are EM. 

In conclusion, we have shown that a parametric interaction between electrostatic 
LOR modes excited by a VLF antenna and FM modes excited by an ELF antenna leads to the non-linear 
excitation of electromagnetic whistler waves.  While a VLF antenna alone also excites EM whistler 
waves, the parametric antenna non-linearly pumps more power (measured in Watts) into the EM whistler 
mode compared with a VLF antenna alone. We find evidence for the existance of EM whistler modes in
$k$-space and real space. Furthermore, we have also investigated the non-linear excitation
of wave modes in frequency space at $\omega_{VLF} \pm \omega_{ELF}$ at different spatial locations in the
simulation domain. We indeed find evidence that the combination frequency is excited, as expected
from theory \cite{FKS87}. This antenna could be used, for example, to generate EM whistler waves 
from a remote antenna to pitch angle scatter high energy particles away from 
satellites which may be subjected to such a harsh evironment.

\subsection{}
\subsubsection{}

\end{document}